\documentclass{ifacconf}
\usepackage{graphicx} 
\usepackage{natbib}
\usepackage{multirow}
\usepackage{color}
\usepackage{amsfonts}
\usepackage{amsfonts}
\usepackage{textcomp}
\usepackage{float}
\usepackage{graphicx}
\usepackage{bm}
\usepackage{booktabs}
\usepackage{amsmath}
\usepackage{amssymb}
\usepackage{makecell}
\usepackage{xfrac}
\usepackage[ruled,vlined,algo2e]{algorithm2e}
\usepackage{algorithm}
\usepackage{algorithmicx}

\usepackage{xcolor}

\begin{document}
\begin{frontmatter}

\title{Data-Driven Koopman-Enhanced Extremum Seeking for Oscillation Damping in Nonlinear Systems}


\thanks[footnoteinfo]{This research was supported by the E-COMP initiative at Pacific Northwest National Laboratory (PNNL).  PNNL is a multi-program national laboratory operated for the U.S. Department of Energy (DOE) by Battelle Memorial Institute under Contract No. DE-AC05-76RL01830.}

\author[First]{Timothy I. Salsbury} 
\author[First]{Min Gyung Yu} 
\author[First]{Sayak Mukherjee} 

\address[First]{Pacific Northwest National Laboratory, 902 Battelle Blvd, Richland WA 99354 USA (e-mail: \{timothy.salsbury; mingyung.yu; sayak.mukherjee\}@pnnl.gov)}

\begin{abstract}                
We propose a novel extremum seeking control (ESC) method that operates in a lifted Koopman state space to minimize the filtered RMS energy in the dominant subspace. The lifted representation provides linear embeddings of nonlinear dynamics, enabling more accurate gradient estimation and dampening of state interference for more consistent ESC performance. Applied to a parameterized, forced, and time-varying Van der Pol oscillator, we show that the approach yields faster and more robust performance than operating ESC on the measured states. These advantages position the method for a diverse range of applications including vibration suppression, motion control, and subsynchronous oscillation mitigation in inverter-dominated power systems.
\end{abstract}

\begin{keyword}
extremum-seeking control, Koopman operator, data-driven control, oscillation damping, nonlinear systems, power systems
\end{keyword}

\end{frontmatter}

\section{Introduction}
Oscillations and instability remain critical challenges in many dynamic system control problems, such as modern power systems, where growing or sustained limit cycles risk equipment damage, system tripping, or blackouts \citep{ping2021deep,zhou2024robust}. Failures in these systems have high societal impacts, with recent examples including the 2017 South Australia blackout and subsynchronous oscillations (SSO) events in Texas wind farms and high voltage direct current (HVDC)-connected systems. Traditional linear control methods often fail to fully mitigate these nonlinear, parameter-dependent phenomena, especially under varying operating conditions or actuator constraints. \cite{mukherjee2025policy} presents a data-driven methodology that tunes the internal control gains of inverters to damp sub-synchronous oscillations.

Control and optimization techniques offer promising mitigation strategies. Model predictive control (MPC) and adaptive damping controllers have been applied to suppress oscillations \citep{korda2018linear}, while data-driven approaches enable real-time control without precise models \citep{mukherjee2021scalable}. Extremum seeking control (ESC) has emerged as a powerful model-free method for online optimization, including limit cycle amplitude reduction \citep{wang2000extremum}. However, conventional ESC performance can be limited by multi-harmonic or non-stationary behavior common in power systems.

The Koopman operator theory provides a data-driven framework to embed nonlinear dynamics into a (approximately) linear space, enabling powerful tools such as spectral analysis and linear control design \citep{korda2018linear}. Recent advances have applied Koopman methods to power grid problems, including robust MPC \citep{korda2018power,zhou2024robust}, deep-learning-based approximations \citep{ping2021deep}, and graph neural network-based lifting \citep{mukherjee2022learning}. Recent works have also investigated Koopman-based designs for closed-loop control with linear quadratic regulators (LQRs) and MPC \citep{caldarelli2025linear, zhou2025deep, shang2025exponential}. Despite these successes, Koopman representations have seen limited integration with real-time, model-free optimization techniques such as ESC \citep{zhao2022global}. This gap is notable, as combining Koopman lifting with ESC could leverage linear embeddings for more reliable gradient estimation and mode-targeted cost functions to reduce signal interference.

This paper proposes a novel extension: performing extremum seeking control directly in a lifted Koopman state space, using a target-free energy-based cost to minimize limit cycle amplitude. We demonstrate the proposed concept on a parameterized, forced, and time-varying Van der Pol (VdP) oscillator that includes a slow drift in the damping coefficient and an external sinusoidal forcing term. This system has some interesting analogies with modern power systems. We show that operating the ESC on the lifted Koopman space improves convergence and robustness compared to operating the ESC directly on the measured states. The method offers advantages in handling strong nonlinearity, multi-harmonic behavior, and measurement noise/interference, making it well-suited for oscillation damping in dynamic systems, including power grids and mechanical vibration control.

The Van der Pol (VdP) oscillator has some interesting analogies with modern power systems such as having inherent negative damping for small amplitudes (energy injection) and positive damping for large, leading to a stable limit cycle. This mirrors self-excited oscillations in power systems where instabilities arise from control interactions (e.g., PLL in inverters providing ``negative damping" at subsynchronous frequencies). In power electronics-dominated systems, sub-synchronous resonance (SSR)/sub-synchronous control interactions (SSCI) often behave like self-excited growing oscillations that stabilize at a limit cycle, which, if unchecked, are exactly like VdP oscillators.

The remainder of the paper is organized as follows: Section~\ref{sec:method} outlines the proposed method; Section~\ref{sec:system} describes the test system including the cost function and the test conditions; and Section~\ref{sec:conclusions} presents conclusions.

\section{Proposed Method}
\label{sec:method}
\begin{figure}[t!]
    \centering
\includegraphics[width=0.4\textwidth]{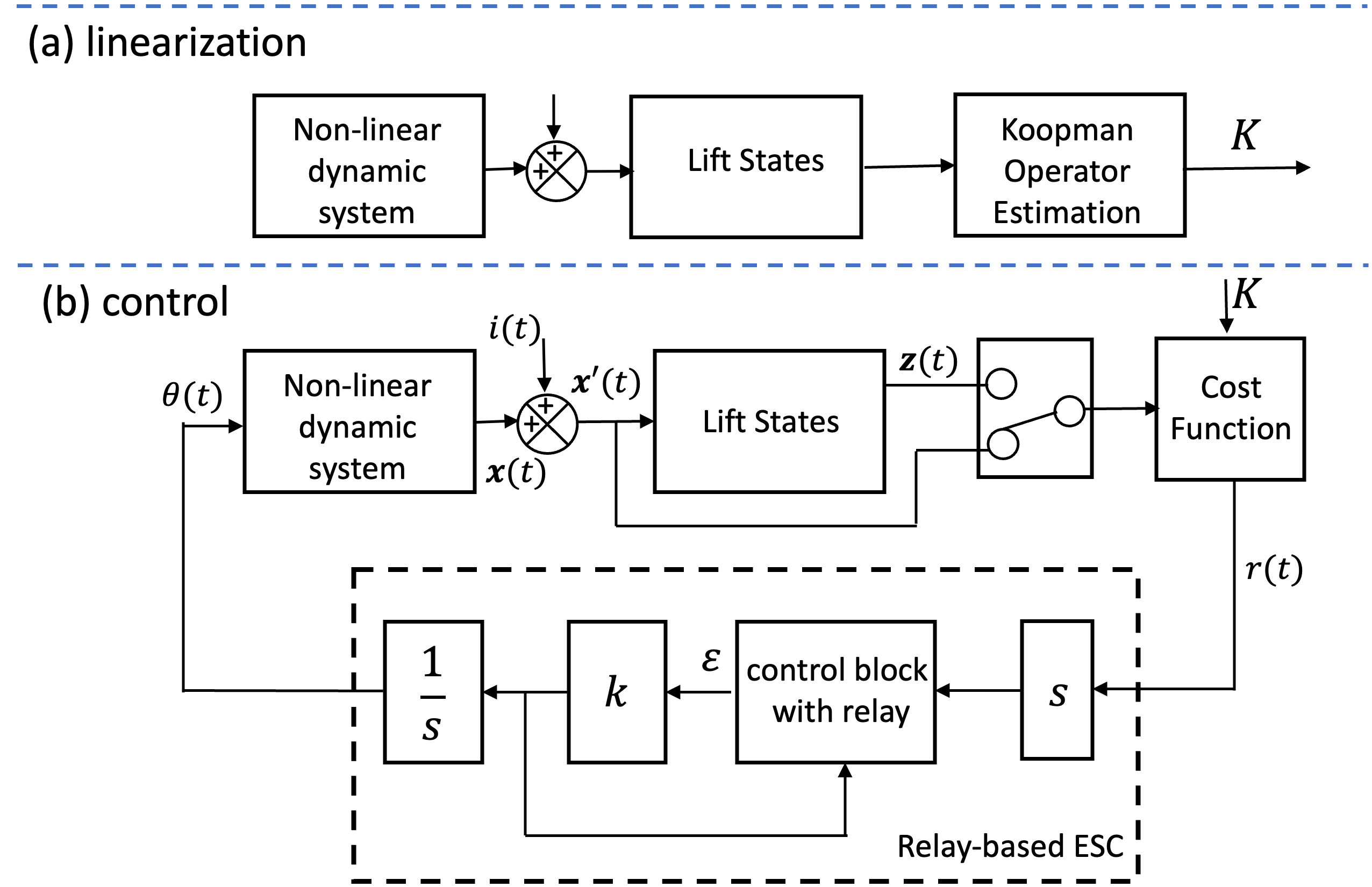}
    \caption{Block diagram of the proposed Koopman-based ESC controller.}
    \label{fig:esc_block}
\vspace{-0.2cm}
\end{figure}

\vspace{-0.2cm}
The method proposed in this paper is illustrated in Figure~\ref{fig:esc_block}. There are two steps required. First, we consider a discrete-time autonomous nonlinear system
\begin{equation}
x_{k+1} = f(x_k), x_0 = x^0,
\end{equation}
where
$
x_k \in \mathbb{R}^n $ denotes the state vector, and $
f : \mathbb{R}^n \to \mathbb{R}^n$ is the nonlinear vector field.
We consider the standard assumptions on Lipschitz continuity:

\textit{Regularity:} $f(x_k)$ is locally Lipschitz, ensuring existence and uniqueness of solutions.


\textit{Forward completeness:} For any initial condition $x_0$, the state sequence $\{x_k\}$ is well-defined for all $k$. Let $\Phi^k$ denote the 
k--fold composition of $f$ 
, i.e.,
\begin{equation}
x_k = \Phi^k(x_0), \quad x_0 = x^0.
\end{equation}

\subsection{Koopman-based Lifted Representation}
\vspace{-0.2cm}
We describe the key characterizations of this representation as follows. The first step is (a) to acquire data that span a broad range of normal operating conditions and use these data to estimate Koopman operators in \(K\). This step also yields the functions for generating lifted states \(z_k\). Let $g : \mathbb{R}^n \to \mathbb{R}^p$ denote a vector of observable functions, and define the lifted state
\begin{equation}
z_k = g(x_k), \quad z_k \in \mathbb{R}^p, \; p \gg n.
\end{equation}

The Koopman operator $\mathcal{K}$ is applied on observables as
\begin{equation}
(\mathcal{K} g)(x_k) = g(f(x_k)) = g(x_{k+1}).
\end{equation} 
When the chosen observables span an invariant subspace under $\mathcal{K}$, the system evolution can be represented with linear lifted dynamics:
\begin{equation}
z_{k+1} = K z_k,
\end{equation}
for some finite-dimensional matrix $K \in \mathbb{R}^{p \times p}$.
The second step (b) is then to use the lifted states in conjunction with the Koopman operator matrix to create a (more) linearized cost function than can be obtained from the raw measured states. The ESC controller is then able to use the linearized cost function as an alternative to improve performance. The block diagram shows a switch selector allowing the control stage to use either the linearized or raw-state derived cost function.

\textit{Main Hypothesis:} The rationale for this approach is that we postulate applying ESC to the lifted space will enable the calculation of a cost function that is more targeted on oscillatory modes of interest, thereby serving to reduce and dampen state measurement interference and also improve the convexity of the cost function, making gradient estimation smoother and more accurate.

\subsection{Relay-based Extremum-Seeking Control}
\vspace{-0.2cm}

ESC is a well-established feedback-driven approach that operates on the ``perturb and observe'' principle \citep{krstic2000stability}. There are many variants of the ESC algorithm and here we use a single-input, single-output (SISO) relay-based ESC strategy as illustrated in Figure~\ref{fig:esc_block}. The relay ESC is easier to tune than traditional demodulation versions as it eliminates the need for scale information of the cost function (see \citep{salsbury2023}). Small updates are applied to a tunable parameter \(\theta(t)\) based on the sign of changes in the filtered cost. The relay logic is implemented as follows. At each discrete time step \(k\), the change in the cost is first passed through a high-pass filter:
\begin{equation}
    \text{hp}(t) = \alpha \cdot \text{hp}(t-\Delta t) + (1 - \alpha) \bigl( r(t) - r(t-\Delta t) \bigr)
\end{equation}
where \(\alpha = e^{-\Delta t / \tau_f}\) is the discrete filter coefficient with time constant \(\tau_f\).

A sign variable \(\epsilon \in \{-1, +1\}\) is maintained and flipped whenever the filtered change is positive and a minimum dwell time \(d_{\rm lim}\) has elapsed since the last flip:
\begin{equation}
    \text{if } \text{hp}(t) > 0 \text{ and } \Delta t_{\rm dwell} \ge d_{\rm lim},
    \quad \epsilon \leftarrow -\epsilon, \quad \Delta t_{\rm dwell} \leftarrow 0.
\end{equation}

The control parameter is then updated using a fixed step size \(k > 0\):
\begin{equation}
    \theta(t) = \theta(t-\Delta t) - k \cdot \epsilon,
\end{equation}
with hard saturation limits \(\theta \in [-5, 5]\) to prevent actuator saturation. This ESC algorithm is applied identically to both the raw-state cost and the lifted Koopman cost, enabling a direct comparison of their performance under measurement interference.

\section{Test System}
\label{sec:system}
A parameterized version of the classic Van der Pol (VdP) oscillator \emph{with time-varying damping and external forcing} is used here:
\begin{equation}
\ddot{x} + \epsilon_0(t)\Bigl[(x-x_0)^2-1-(\theta-\theta^*)^2\Bigr]\dot{x}+\mu^2(x-x_0)=f(t)
\label{eq:vdp}
\end{equation}
where the damping coefficient is time-varying according to
\begin{equation}
\epsilon_0(t)=\epsilon_0+\sin(0.005\,t)
\end{equation}
and the external forcing term is
\begin{equation}
f(t)=2.2\sin(4.5\,t).
\end{equation}

A VdP test system was also used in Wang and Krstić (2000) to demonstrate the ability of ESC to minimize oscillation amplitude. As in that example, \(\theta-\theta^*\) directly controls the amplitude of the limit cycle while \(x_0\) sets the offset of the oscillation. The parameter \(\mu\) sets the nominal frequency of the limit cycle, and the base damping \(\epsilon_0\) controls the speed of the transient. The assumption is that \(\theta^*\) is constant and unknown, while \(\theta\) is the available control input. The system remains in a limit cycle for all admissible \(\theta\) values and emulates realistic scenarios in which the controller can only reduce (but not eliminate) the oscillation amplitude.

The equivalent first-order state-space form (with state vector \([x(t),\;y(t)]^\top\), where \(y(t)=\dot{x}(t)\)) is
\begin{align}
\dot{x} &= y, \label{eq:state1} \\
\dot{y} &= -\epsilon_0(t)\Bigl[(x-x_0)^2-1-(\theta-\theta^*)^2\Bigr]y \nonumber\\
&\quad -\mu^2(x-x_0)+f(t). \label{eq:state2}
\end{align}

The continuous-time model is discretized with a fixed step \(\Delta t=0.005\,\text{s}\) for both simulation and control.

\subsection{Interference Model}
To simulate realistic sensor corruption and higher-frequency disturbances (e.g., harmonic pollution or external vibrations), we introduce additive sinusoidal interference to the measured state vector. This interference is defined as
\begin{equation}
    \mathbf{i}(t) = A_1 \sin(3.2\, t) + A_2 \sin(7.8\, t),
\end{equation}
where \(A_1 = 10.5\) and \(A_2 = 11.1\).

The measured state at each time step is then obtained by
\begin{equation}
    \mathbf{x}'(t) = \mathbf{x}(t) + \mathbf{i}(t),
\end{equation}
where $\mathbf{x}(t)$ is the true state of the Van der Pol oscillator. This corruption is applied in two places to ensure consistency between training and operation:
\begin{enumerate}
    \item \textbf{During Koopman operator training}:  
    The data for estimating Koopman operators are constructed from corrupted states, so the learned operator matrix \(K\) is robust to the expected interference.
    \item \textbf{During ESC control}:  
    The controller only has access to the corrupted measurement \(\mathbf{x}'(t)\). This measured state is used to compute both the raw-state cost and the lifted Koopman cost.
\end{enumerate}

This interference model creates a challenging scenario in which the raw-state cost becomes inaccurate due to unwanted frequencies, while the lifted Koopman approach can selectively focus on the dominant modes of the true dynamics, thereby demonstrating a structural robustness advantage.

\subsection{Koopman Lifted Space}
\label{sect:koopman}
We lift the states $[x,y]$ to $z$ using polynomials (degree 3), i.e.,
\begin{align}
\label{eq:basisF}
    z = [1,x,y,x^2,xy,y^2,x^3,x^2y,xy^2,y^3]^T
\end{align}
This projects the original states into a linear 10-dimensional lifted space. We generate these data by simulating trajectories from the test system at various randomly selected initial conditions and parameters.

We simulate the VdP system to collect pairs of consecutive states, and in our discrete-time execution of the simulation, this corresponds to gathering $M$ snapshot pairs of current states $X = \begin{bmatrix} x_1 & x_2 & \cdots & x_M \end{bmatrix}$ and next states $Y = \begin{bmatrix} x_2 & x_3 & \cdots & x_{M+1} \end{bmatrix}$. For the VdP system, we generate 100 simulations with $\theta \in [-5,5]$ yielding data size $X,Y \in \mathbb{R}^{n\times M}$ where $n$ is the state dimension and $M$ is large from the multiple discrete points from each trajectory and multiple $\theta$ values.

We then apply the set of basis functions $g:\mathbb{R}^{n} \longrightarrow \mathbb{R}^{p}$ with $p>n$ and based on the above proposed basis functions $p=10$
\begin{align}
\label{eq:psiX}
\Psi_X & = \begin{bmatrix} g(x_1) & g(x_2) & \cdots & g(x_M) \end{bmatrix} \in \mathbb{R}^{p \times M},\\
\Psi_Y & = \begin{bmatrix} g(y_1) & g(y_2) & \cdots & g(y_M) \end{bmatrix} \in \mathbb{R}^{p \times M}
\end{align}

Using least-squares regression, we then find the Koopman operator matrix $K$ that minimizes the residual, i.e.,
\begin{align}
    K = \arg\min_{K} \|\Psi_Y - K \Psi_X\|_F^2 = \Psi_Y \Psi_X^\dagger
\end{align}

\subsection{Projection onto Dominant Koopman Modes}
The measured (corrupted) state \(x'_k\) is first mapped to the observable vector
\[
z_k = g[x'_k] \in \mathbb{R}^p,
\]
where \(g(\cdot)\) is the polynomial lifting function (Eq.~\eqref{eq:basisF}) and \(p=10\). Instead of selecting modes solely by eigenvalue magnitude near the unit circle, we adopt an {\em energy-based criterion} that directly reflects the contribution of each mode to the observed data. Specifically, the modal coefficients on the training data are obtained as:
\[
\mathbf{C} = V^{+} \Psi_X \in \mathbb{R}^{p \times M},
\]
where \(V\) contains the right eigenvectors of the learned Koopman operator matrix \(K\) and \(\Psi_X\) is the matrix of lifted training snapshots (Eq.~\eqref{eq:psiX}). The average energy of mode \(j\) is then
\[
E_j = \frac{1}{M} \sum_{k=1}^{M} |c_j(k)|^2.
\]
The \(N\) modes with the largest \(E_j\) are retained, forming the dominant-mode matrix \(V_{\rm dom} \in \mathbb{R}^{p \times N}\) whose columns are the corresponding right eigenvectors of \(K\).  
The projection onto this subspace is
\[
\mathbf{c}_{\rm dom,k} = V_{\rm dom}^{+} z_k,
\]
and the instantaneous energy used by the ESC is
\[
y_{\rm out,k} = \|\mathbf{c}_{\rm dom,k}\|_2^2.
\]
For the system considered, \(N=8\) was found to yield the smoothest convex cost landscape and the strongest closed-loop performance while still attenuating the added sinusoidal interference. A data-driven ranking of Koopman modes by their average energy contribution $E_j$ is a common and effective approach in the literature (see, e.g., \citep{kou2017improved}; \citep{hua2017high}). This criterion selects the modes that are most observable in the training data and therefore most relevant for the cost function, as opposed to purely spectral criteria based on eigenvalue magnitude near the unit circle.

\subsection{Cost Function}
\vspace{-0.2cm}
We calculate a cost function for the ESC following the same approach used in Wang and Krstic (2000). We denote the cost function by \(r_k\), and it is computed in two stages: first, an instantaneous energy measure \(y_{\rm out,k}\), then a standard amplitude detector consisting of high-pass and low-pass filtering. The instantaneous energy \(y_{\rm out,k}\) is obtained differently for the raw-state and lifted cases:

\begin{itemize}
\item \textbf{Raw-state cost:}
\[
y_{\rm out,k} = \frac{1}{\sqrt{2}} \|x'_k\|_2
\]

\item \textbf{Lifted Koopman cost:} First, lift the measured state to the observable vector \(z_k = g[x'_k]\). Then project onto the dominant Koopman modes:
\[
\mathbf{c}_{\rm dom,k} = V_{\rm dom}^{+} z_k,
\]
and compute
\[
y_{\rm out,k} = \|\mathbf{c}_{\rm dom,k}\|_2^2 = \sum_j |c_j(k)|^2.
\]
\end{itemize}

The final cost signal \(r_k\) is obtained by passing \(y_{\rm out,k}\) through the classical amplitude detector (high-pass filtering followed by squaring and low-pass filtering):
\[
y_{\rm hp,k} = \alpha_{\rm hp} y_{\rm hp,k-1} + (1 - \alpha_{\rm hp})(y_{\rm out,k} - y_{\rm out,k-1}),
\]
\[
y_{\rm lp,k} = \alpha_{\rm lp} y_{\rm lp,k-1} + (1 - \alpha_{\rm lp}) y_{\rm hp,k}^2,
\]
\[
r_k = \max(y_{\rm lp,k}, 0),
\]
where \(\alpha_{\rm hp} = e^{-\omega_h \Delta t}\) and \(\alpha_{\rm lp} = e^{-\omega_l \Delta t}\) are the discrete filter coefficients. This \(r_k\) therefore represents the low-pass filtered RMS amplitude of either the raw measured state or the energy contained in the dominant Koopman modes. The ESC then minimizes \(r_k\) by adjusting the control input \(\theta_k\).
\vspace{-0.2cm}
\subsection{Summary of parameters}
\vspace{-0.2cm}
All parameters used in the VdP model, the interference model, and in the ESC algorithm are summarized in Table~\ref{tab:vdp_parameters}.
\begin{table}[ht]
\centering
\small
\caption{Simulation and algorithm parameters for the Van der Pol system}
\label{tab:vdp_parameters}
\begin{tabular}{l l}
\hline
\textbf{Parameter} & \textbf{Value} \\
\hline
Base \(\epsilon_0\) & 1.0 \\
Drift in \(\epsilon_0(t)\) & \(+1 \sin(0.005 t)\) \\
Forcing term & \(2.2 \sin(4.5 t)\) \\
\(\mu\) & 1.0 \\
\(x_0\) & 6.0 \\
\(\theta^*\) & -3.0 \\
\hline
Interference amplitudes & \(A_1=10.5\), \(A_2=11.1\) \\
Interference frequencies & 3.2 rad/s, 7.8 rad/s \\
\hline
Relay step size \(k\) & \(\Delta t / 15\) \\
Dwell time $d_{\rm lim}$ & 15 s \\
Saturation limits on \(\theta\) & \([-5, 5]\) \\
Initial \(\theta(0)\) & 2.0 \\
\hline
Koopman training trajectories & 100 \\
\hline
\end{tabular}
\end{table}

\subsection{Performance Metrics}
\vspace{-0.2cm}
To evaluate the performance of the proposed method, we consider both the control input $\theta(t)$ and the corresponding output response $r(t)$. The following metrics are used:
\begin{itemize}
    \item \textbf{Tracking performance:} The overall tracking performance is quantified using the integral error metrics
    \begin{align}
    \mathrm{IAE}_\theta &= \int_0^T |\theta(t) - \theta^*| \, dt, \\
    \mathrm{ISE}_\theta &= \int_0^T (\theta(t) - \theta^*)^2 \, dt.
    \end{align}
    These metrics capture both transient and steady-state behavior. The IAE provides an interpretable measure of accumulated error, while the ISE penalizes large deviations more strongly.

    \item \textbf{Convergence speed:} The convergence speed is evaluated using the first hitting time
    \begin{equation}
    T_\theta = \min \left\{ t : |\theta(t) - \theta^*| < \epsilon \right\},
    \end{equation}
    where $\epsilon$ is a predefined tolerance. This metric reflects how quickly the controller reaches the neighborhood of the optimum.


    \item \textbf{Steady-state output performance:} 
    To evaluate the output behavior, a range-normalized steady-state error is considered:
    \begin{equation}
    e_r^{\mathrm{ss}} = \frac{|\bar{r} - r^*|}{r_{\max} - r_{\min}},
    \end{equation}
    where $\bar{r}$ denotes the average of $r(t)$ over the final portion of the trajectory, and $r^*$ is the optimal output value obtained from the static map.
The output is normalized by its range to ensure fair comparison across different scales. The steady-state value $\bar{r}$ is computed over the last $30\%$ of the data to exclude transients.
\end{itemize}
\section{Experimental Results}
\vspace{-0.2cm}
For the experimental tests, we first generated 100 training trajectories using the parametrized VdP system and collected the state time series data as described in Section~\ref{sect:koopman}. For the training data, the $\theta$ value was varied randomly between -5 and +5. These data were then used to determine the Koopman operator matrix $K$. Table~\ref{tab:mode_selection} lists the 10 modes and their eigenvalues showing which ones were retained based on the energy-based criterion \(E_j\) in the training data. Retaining the 8 most energetic modes produced the smoothest convex static map and the fastest ESC convergence.

\begin{figure}[ht]
\centering
\includegraphics[width=3.25in]{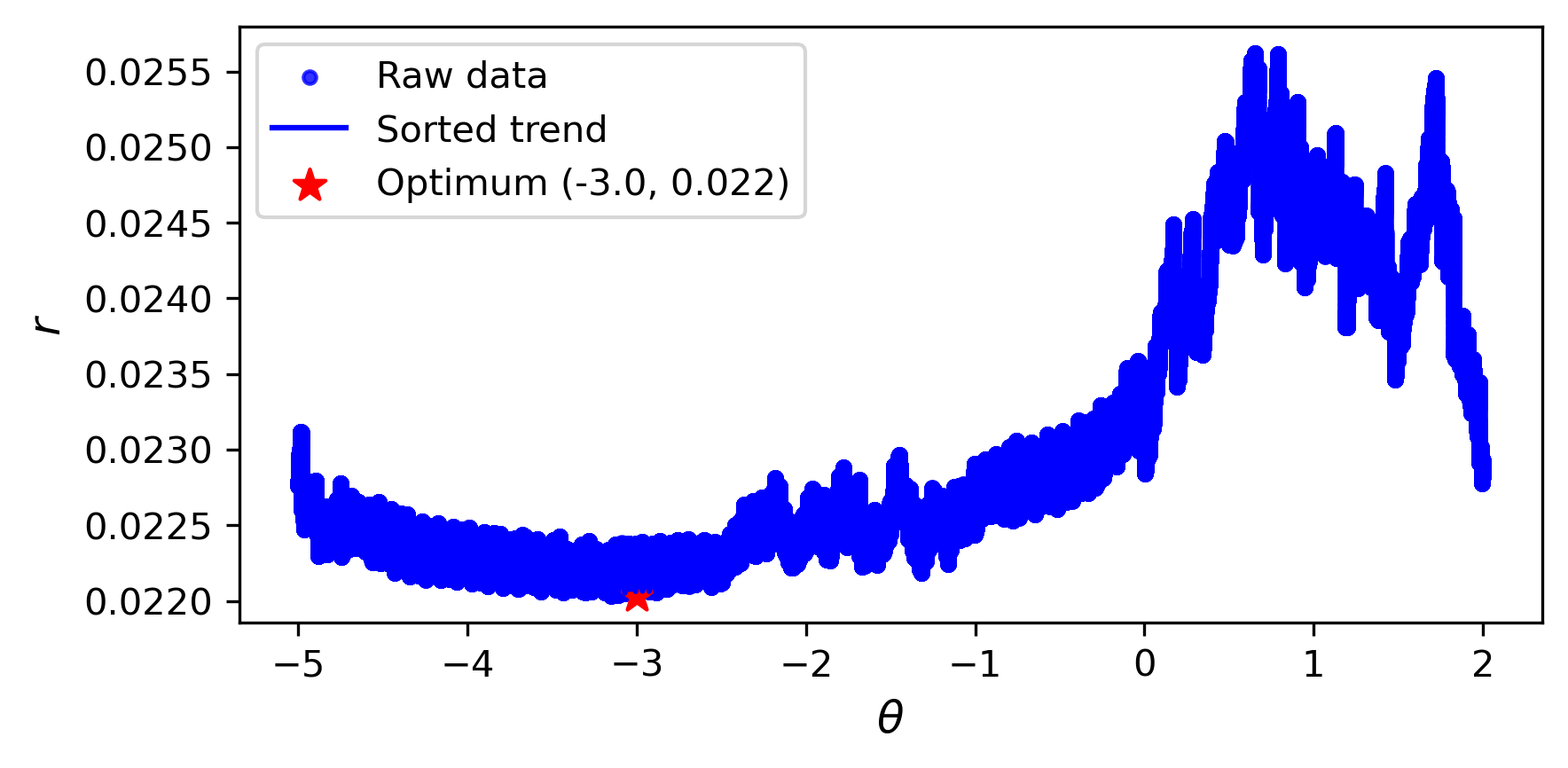}
\vspace{-0.3cm}
\caption{Static map in the non-lifted space $r$ vs $\theta$}
\vspace{-0.2cm}
\label{fig:static_original}
\end{figure}

\begin{figure}[ht]
\centering
\includegraphics[width=3.25in]{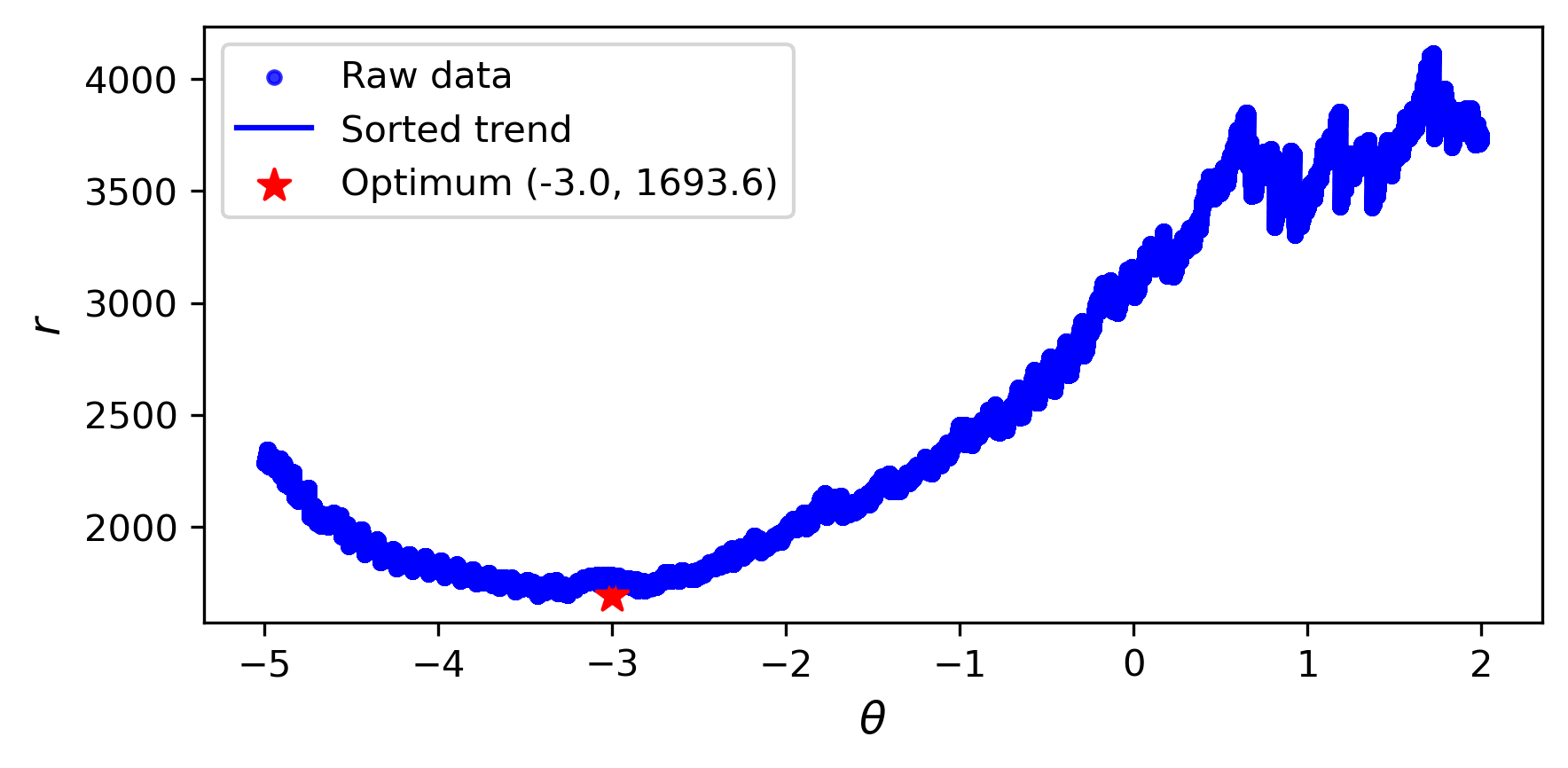}
\vspace{-0.3cm}
\caption{Static map in the lifted space $r$ vs $\theta$}
\vspace{-0.1cm}
\label{fig:static_lifted}
\end{figure}

\begin{table}[t]
\centering
\caption{Summary of selected modes and their eigenvalues}
\label{tab:mode_selection}
\begin{tabular}{cccr}
\toprule
eigenvalue \(\lambda\) & \(|\lambda|\) & Selected \\
\midrule
\(1.000000 + 0.000000j\) & 1.000000 & No \\
\(0.999290 + 0.000000j\) & 0.999290 & Yes \\
\(0.999109 + 0.000000j\) & 0.999109 & Yes \\
\(0.998697 + 0.000000j\) & 0.998697 & No \\
\(0.989538 + 0.000000j\) & 0.989538 & Yes \\
\(0.937694 + 0.000000j\) & 0.937694 & Yes \\
\(0.896287 + 0.000000j\) & 0.896287 & Yes \\
\(0.848728 + 0.000000j\) & 0.848728 & Yes \\
\(0.734663 + 0.105522j\) & 0.742203 & Yes \\
\(0.734663 - 0.105522j\) & 0.742203 & Yes \\
\bottomrule
\end{tabular}
\end{table}
To first understand the optimization landscape, we construct static maps as shown in Figures~\ref{fig:static_original}--\ref{fig:static_lifted}. In both cases, the horizontal axis represents $\theta$, and the vertical axis represents the cost signal $r$. The data are obtained from a ramp experiment in which $\theta$ is gradually decreased from $2$ to $-5$, while recording the corresponding $r$ values. At the beginning of the sweep (near $\theta=2$), both systems show noticeable oscillations, which gradually diminish as $\theta$ decreases.

Figure~\ref{fig:static_original} shows the relationship between $\theta$ and $r$ in the non-lifted space. The map appears highly noisy, with significant variations in $r$ observed even at similar $\theta$ values. This indicates that the system response is not uniquely determined by $\theta$, making the underlying structure difficult to interpret. Although the optimum ($\theta^* = -3$) can be identified, the lack of a clear convex shape complicates the optimization process. In particular, such irregularity makes it challenging for gradient-based optimization methods, like ESC, to reliably track the optimal operating point.

Figure~\ref{fig:static_lifted} presents the static map in the lifted space. Although the scale of the vertical axis differs due to the lifting transformation, we need to focus on the overall shape of the map rather than the absolute magnitude. Compared to the non-lifted case, the lifted map shows a clear convex structure with a well-defined minimum. Although some residual distortion remains around the optimum, the lifted map exhibits markedly improved convexity and a clearer alignment with the theoretical optimum $\theta^* = -3$. The slight shift in the optimal point is expected, as the lifting transformation modifies the representation of the system dynamics. Nevertheless, this discrepancy is small and does not affect the qualitative behavior. More importantly, the lifted representation significantly improves the smoothness and reliability of the static map. This enhanced structure provides a more favorable landscape for ESC to converge more efficiently and operate more reliably around the optimal point.

\begin{figure}[ht]
\centering
\includegraphics[width=3.25in]{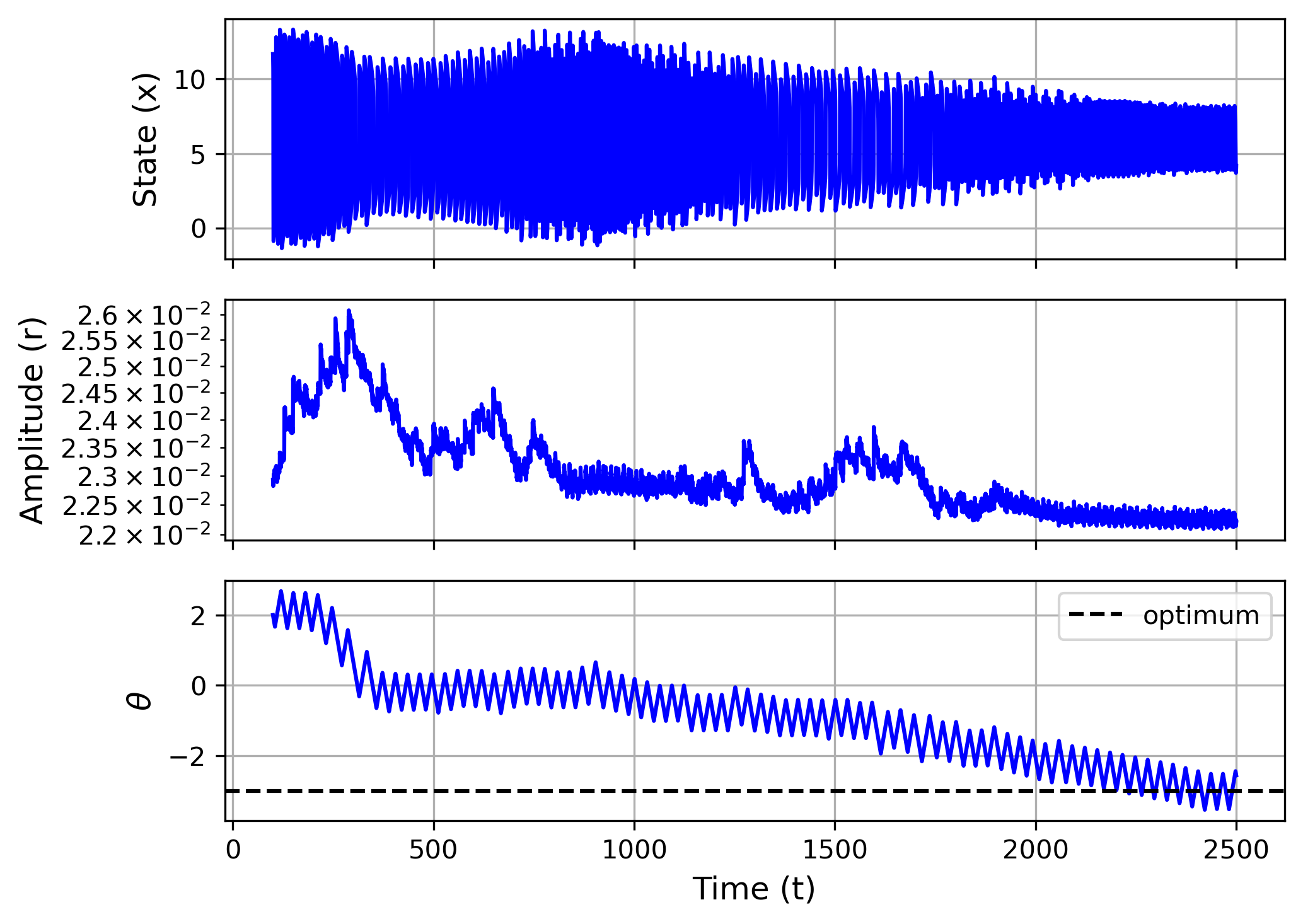}
\vspace{-0.3cm}
\caption{ESC response in the non-lifted space}
\vspace{-0.2cm}
\label{fig:esc_original}
\end{figure}

\begin{figure}[ht]
\centering
\includegraphics[width=3.25in]{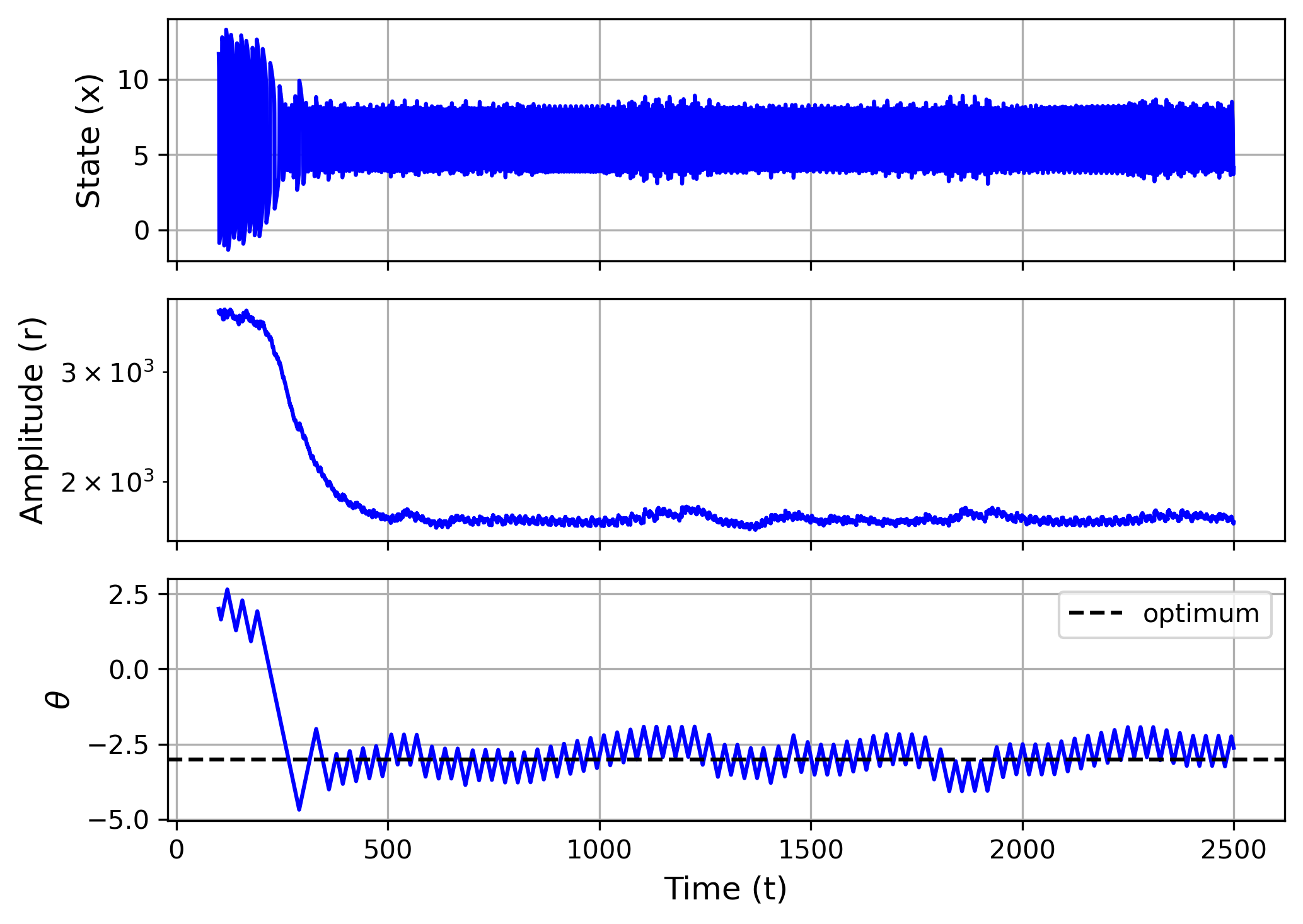}
\vspace{-0.3cm}
\caption{ESC response in the lifted space}
\vspace{-0.2cm}
\label{fig:esc_lifted}
\end{figure}

We next evaluate the performance of extremum seeking control (ESC) in both the non-lifted and lifted spaces, as shown in Figures~\ref{fig:esc_original}--\ref{fig:esc_lifted}. The plots are shown for $t \in [100, 2500]~{\rm secs}$, since data logging is initiated after $t=100~{\rm secs}$ to exclude the initial transient dynamics associated with the system initialization in the ESC scheme. In each figure, the first subplot shows the system state, the second subplot shows the $r$, and the third subplot shows the parameter estimate $\theta$. The optimal value of $\theta^*$ is indicated by a dashed horizontal line in the third subplot.

Figure~\ref{fig:esc_original} illustrates the ESC behavior in the non-lifted space. The convergence toward the optimal point is notably slow, requiring approximately $t \approx 1977~{\rm secs}$ to reach the vicinity of the optimum. The parameter $\theta$ evolves gradually, and this reflects the difficulty of navigating the irregular optimization landscape observed in the static map. Correspondingly, the $r$ exhibits slow improvement and remains relatively noisy over time. The system state also shows sustained oscillatory behavior with slowly decaying amplitude.

In contrast, Figure~\ref{fig:esc_lifted} shows the ESC response in the lifted space. The controller rapidly drives the parameter estimate toward the optimal point, reaching the vicinity of the optimum at approximately $t \approx 258~{\rm secs}$. A transient adjustment is observed around $t \approx 220~{\rm secs}$, where $\theta$ shows a sharp drop before settling near the optimal value. After this transient phase, the parameter estimate remains tightly regulated around the optimum.
This improved convergence behavior is also reflected in the system state and $r$. The state represents significantly faster damping of oscillations and quickly settles into a stable regime. Similarly, the $r$ decreases rapidly and maintains a smooth and stable profile after convergence, in contrast to the slower and noisier behavior observed in the non-lifted case.

\begin{table}[t]
\centering
\caption{Performance comparison between non-lifted and lifted ESC}
\label{tab:performance}
\begin{tabular}{lcccc}
\hline
Method & $\mathrm{IAE}_\theta$ & $\mathrm{ISE}_\theta$ & $T_\theta$ & $e_r^{\mathrm{ss}}$ \\
\hline
Non-lifted & 5125.73 & 14720.91 & 1977 & 0.08816 \\
Lifted     & 1445.30 & 3166.17  & 258  & 0.02486 \\
\hline
\end{tabular}
\end{table}

Table~\ref{tab:performance} demonstrates that the lifted ESC significantly improves performance across all metrics. In terms of tracking accuracy, both $\mathrm{IAE}_\theta$ and $\mathrm{ISE}_\theta$ are substantially reduced, with approximately $3.5\times$ lower IAE and $4.6\times$ lower ISE compared to the non-lifted case. The convergence speed is significantly improved, with the lifted method reaching the optimum approximately $7.7\times$ faster. In addition, the normalized steady-state cost error $e_r^{\mathrm{ss}}$ is reduced by about $3.5\times$, indicating closer convergence to the optimal value.

Overall, the comparison clearly demonstrates the advantage of performing ESC in the lifted space. The lifted representation reshapes the underlying optimization landscape into a more convex and well-structured form, which directly enhances the effectiveness of the controller. As a result, ESC achieves substantially faster convergence, improved stability in the system state, and more consistent operation around the optimal point.

\section{Conclusions}
\label{sec:conclusions}
In this paper, we proposed combining a Koopman-based state transformation to create an improved cost function for extremum-seeking control. The approach allows the lifted states to be projected onto oscillatory modes of interest, thereby serving to reduce and dampen state measurement interference and also improve the convexity of the cost function, making gradient estimation smoother and more accurate. The approach is data-driven and model-free, with only minimal configuration parameters required for deployment. We used a simple VdP model to illustrate the utility of the approach and demonstrated significantly improved performance from the ESC when operating on the cost function derived from the lifted states. We suggest that using Koopman-based lifted states in conjunction with a real-time optimizer like ESC could have wide applicability across a range of applications, including, among others, online tuning of power system stabilizers for subsynchronous and inter-area oscillation damping, adaptive vibration suppression in mechanical structures and wind turbines, flutter control in aerospace systems, and limit-cycle mitigation in robotics and biomedical devices. By selectively minimizing energy in the dominant Koopman modes, the approach described in the paper provides a practical, model-free pathway to enhance stability in uncertain nonlinear environments.

\bibliography{refs}             
                                                             
\end{document}